\documentclass{article}

\usepackage{graphicx}
\usepackage{setspace}
\usepackage{amsmath}
\usepackage{todo}
\usepackage{url}
\usepackage{cite}
\usepackage{hyperref}

\title{Self-Reproduction and Evolution in Cellular Automata: 25 Years after Evoloops}
\author{Hiroki Sayama$^{1,2}$ and Chrystopher L. Nehaniv$^{3}$\\~\\
$^1$ Binghamton Center of Complex Systems,\\ Binghamton University, State University of New York,\\ Binghamton, NY, USA\\
$^2$ Waseda Innovation Lab, Waseda University, Tokyo, Japan\\
$^3$ Waterloo Institute for Complexity and Innovation,\\
Departments of Systems Design Engineering,\\  of Electrical \& Computer Engineering, \\ and of Applied Mathematics,
 University of Waterloo,\\ Waterloo, Ontario, Canada}
\date{\today}

\begin{document}

\maketitle

\begin{abstract}
The year of 2024 marks the 25th anniversary of the publication of evoloops, an evolutionary variant of Chris Langton's self-reproducing loops which proved constructively that Darwinian evolution of self-reproducing organisms by variation and natural selection is possible within deterministic cellular automata. Over the last few decades, this line of Artificial Life research has since undergone several important developments. Although it experienced a relative dormancy of activities for a while, the recent rise of interest in open-ended evolution and the success of continuous cellular automata models have brought researchers’ attention back to how to make spatio-temporal patterns self-reproduce and evolve within spatially distributed computational media. This article provides a review of the relevant literature on this topic over the past 25 years and highlights the major accomplishments made so far, the challenges being faced, and promising future research directions.
\end{abstract}

\section{Introduction}

Christopher Langton, the founder of Artificial Life as an established research field, is also well known for several important contributions he made to cellular automata research. Probably the most famous of his cellular automata work is his self-replicating loop model \cite{Langton1984} published in 1984, three years before the first Artificial Life conference was held. It is commonly considered the ``third generation'' of self-reproducing cellular automata, after von Neumann's original self-reproducing automata \cite{von Neumann1966} and Codd's simplified version \cite{Codd1968,Hutton2010}, from which Langton's loops were derived. Langton described his ambition in cellular automata-based artificial life research as follows:
\begin{quote}
    ``If we could populate a large area with multiple copies of such reproducing colonies, and {\it introduce variation} into at least the portion of the description that codes for the extra machinery, we would have all of the raw material necessary for natural selection to operate among variants and hence we would have {\it a sufficient basis for the process of evolution}.'' --- Chris Langton \cite{Langton1986} (italics ours)
\end{quote}
 
Creating such a demonstrative evolutionary process within artificial media like cellular automata was one of the original goals set by founding pioneers of artificial life, including John von Neumann \cite{von Neumann1966} and is clearly articulated by Langton \cite{Langton1984,Langton1986} in the quote here. 
This goal was later achieved in the form of evoloops \cite{Sayama1999a,Sayama1999b} published in 1999.
Evoloops were the first cellular automata-based artificial life that exhibited true Darwinian evolution of self-reproducing organisms by spontaneous variation and natural selection.  For conceptual precision in reviewing these developments and focusing on future directions on realizing von Neumman's vision,  we shall distinguish between {\em self-replication} (of identical copies of organisms) and {\em self-reproduction} (in which variation in the copies is both possible and inheritable) in Section~\ref{background}.

The year of 2024 marks the 40th anniversary of Langton's self-replicating loops and the 25th anniversary of self-reproducing evoloops. This line of research has remained, however, somewhat unpopular and unexplored ever since, likely because of the lack of rigorous theories, generalizability of models, and immediate applications to practical problem solving. Nonetheless, several important developments have since been made and, more recently, artificial life researchers have begun to pay attention again to how to make patterns self-replicate and evolve within spatially distributed computational media (as reviewed in this article), potentially leading to open-ended evolution that keeps generating new forms of life indefinitely while exhibiting behavior of increasing complexity without converging toward some optimum \cite{packard2019a,packard2019b,stanley2019,Taylor2016}.

In this review, we aim to summarize the relevant literature on this topic over the last 25 years and highlight the major accomplishments made so far, the challenges being faced, and promising future research directions. We hope this review will serve as an unofficial sequel of the earlier review by Sipper \cite{Sipper1998} that was published about the same time as the discovery of evoloops. Our focus in this article will thus be on the evolutionary self-reproducing automata models.


\section{Background, Key Concepts and Advances}\label{background}

We mentioned above and will later return to the central themes of von Neumann's seminal work on self-reproduction in automata (1948) \cite{vN}. We first highlight the various contributions of Moore, Langton, Sayama, and others, and their impact along the way in this narrative. For a number of years, work in this field focused somewhat over-restrictively on aspects emphasized in Langton's work but now returns to the more general framing and original insights of von Neumann which embrace a broader notion of self-reproduction centered on the capacity for inheritable mutation in the production of offspring.
First we review Moore's criterion for self-replication, distinguish trivial and non-trivial self-replication, draw the distinction between self-replication and self-reproduction, and then discuss the related concept of `individuals' which these definitions tacitly presuppose and which will be important in relating self-reproduction to evolution.\footnote{The term {\em individual} occurs in three intertwined ways in this subject: first, there is the debate in biology of what constitutes an individual as a living self-maintaining organism (see Sec.~\ref{individual-bio}); second, there is the notion of individual as an entity in a population undergoing a process of Darwinian evolution, which, like  the  primitive terms `line' and `point' in axiomatic geometry, may be treated as an undefined primitive entity in any system satisfying Darwin's axioms for an evolutionary process (see Sec.~\ref{individual-evo}); and finally there is the useful proxy notion of an individual in constructive artificial life studies as a configuration in cellular space surrounded by quiescent cells that replicates or reproduces, i.e.,
 its instantiation at a single point in space-time (see Sec.~\ref{individual-CA}). An underlying subtext of this review is to illustrate how the study of self-reproduction and evolution comes to integrate these notions of individual into one.}
We then introduce von Neumann's problems regarding self-reproduction and   complexity, and his solutions or partial solutions to them. 
Then Langton's views are described and then von Neumann's. This leads us to a brief review of evolution as it has been achieved in cellular automata in Section~\ref{sec:evolution} and to future challenges and open problems in Section~\ref{sec:Challenges}.

\subsection{Moore's Criterion for Self-Replication and Beyond}
\label{individual-CA}

To support the study of self-reproduction from a logical perspective,
 E.\ F.\ Moore's Criterion (1962) \cite{Moore1962} uses the notion of a configuration surrounded by empty space in a cellular automaton as proxy for an `individual'.  A configuration is a pattern of state activation within a bounded region of the space.  \\[1ex]

 \noindent
 {\bf Moore's Criterion for Self-Replication}: 
If $C$ is a configuration in an infinite cellular space surrounded by quiescent cells, 
 for all $N\geq 1$ \, then there exists a time $t_N$ at which there exist at least $N$ copies of $C$.\\[1ex]

Lones and Reggia \cite{LohnReggia1997} extended and made this criterion more elaborate; and then, using it to guide an evolutionary algorithm, they evolved self-replicating cellular automata configurations and their local update rules (unexpectedly self-replicating structures were moving!). Nehaniv and Dautenhahn \cite{Nehaniv1998} discuss further extension of Moore's criterion so as to apply to self-reproduction, based on quantitative degree of spatiotemporial matching of configurations that may be non-exact copies.

\subsection{Non-Triviality of Self-Replication}\label{TrivialSelfRep}

There are many cases though that satisfy Moore's criteria or its physical analogues, but 
are borderline cases, not representative of living things, just patterns growing regularly, without discernable `individuals'. 
Their dynamics ensures replication without bound. 

Examples include: (1) the growth of crystals or snowflakes, 
(2) spreading activation\footnote{For example, a binary cellular automata with local states 0 `Off' (quiescent), where a grid cell is unchanged in the next time step unless it has any neighbor that is 1 `On' in which case its state is `On' in the next time step. Although  Moore's criterion is technically violated  here due to lack of surrounding quiescent cells,  similar examples of single active cells  for many elementary 1D cellular automata rules do satisfy the criterion.) }, and (3) the 
modulo prime cellular automata \cite{Amoroso1971} which allow
any 1D or 2D compact configuration to replicate using a special rule, where 
the new state of a cell is the modulo $p$ sum of states of neighbors where $p$ is any prime number and the local states are $\{0,1,\ldots, p-1\}$. 
This results in the reliable replication of any pattern, but this replication is a property of the `physics' of the cellular space rather than dependent on any recognizable activity of individuals or their constitution.

A borderline trivial example of self-replication is the example of {\bf prions} which are proteins taking conformations (shapes), either normal or pathological.  A protein with the same amino acid sequence can fold up in different configurations to produce a typical `healthy' conformation or one which causes disease and death. 
Pathological conformation is transmitted to healthy proteins by interaction with the proteins having the pathological conformation, e.g.\ Creutzfeldt-Jakob disease (vCJD) or `mad cow disease' in humans causing degeneration of the nervous system \cite{cjd}.
If prion proteins were to be considered `individuals', that would be an argument for the single cells of a 1D cellular automaton with spreading activation being `individuals' as well, though they clearly should not be.

\subsection{Self-Replication vs Self-Reproduction}

We have seen Moore's criterion for self-replication and mentioned generalizations of his criterion to non-exact copies above, and we discussed some degenerate examples that do not reflect what one might expect for non-trivial self-replication or self-reproduction. However, the astute reader will observe that we still lack rigorous definitions distinguishing self-replication and self-reproduction. Therefore,
we next articulate a definition for self-replication here and the two main competing criteria for self-reproduction due to Langton and von Neumann, respectively, in Sections~\ref{LangtonCriterion} and \ref{VNcriterion}, and then go on to describe how they arose and more recent developments in the sequel. It will turn out (1) that Langton's criterion accepts some self-replicating systems, such as the Langton loop \cite{Langton1984} or Sayama's SDSR loop \cite{Sayama1998} (see below), although these are excluded  by von Neumann's criterion for self-reproduction as it requires viable inheritable variation, and (2) that Langton's criterion is not broad enough to recognize self-reproduction in systems using self-examination to reproduce (a mechanism proposed by von Neumann -- see Sec.~\ref{vNproblem}). However, both Langton's and von Neumann's criteria do recognize evoloop~\cite{Sayama1999a}, described below, as self-reproducing.

\subsubsection{Non-trivial Self-Replication and Self-Reproduction}

Both von Neumann and Langton were concerned to exclude such trivial cases of self-replication and  introduced different criteria for non-trivial {\em self-reproduction} (Sections~\ref{LangtonCriterion}~\&~\ref{VNcriterion}).
{\em Self-replication} in this review will refer to production of offspring that are exact copies of the parent. This is consistent with Moore's criterion for self-replication. 
Langton created the  first artificial non-trivial implemented self-replicating system in his 1984 paper \cite{Langton1984}.  Von Neumann gave the first rigorous descriptions and analyses of self-reproducing automata \cite{von Neumann1966}.

\subsubsection{Individuals}\label{Individuals}
\label{individual-bio}
A key property of living systems is that they comprise separate {\em individuals} distinct from both their environment and also from other individuals.
The nature of this separation is addressed by studies in the evolution of differentiation in multicellular life, of cancer, and also in  
the conception of living systems as autopoietic unities \cite{Varela1974}, i.e., self-producing entities that create and maintain an identity as an ongoing process of continually regenerating their own organization and components as the activity of such a network.  

What reproduces are individuals.  But what does this really mean?
As J.J.\ von Uexk\"{u}ll \cite{vonUexkull1926} pointed out, individual
living organisms are generally `centrifugal' in character,  with the whole coming first, and its development and differentiation flowing outwards. In contrast, artificial systems (robots, factories, bicycles, etc.) are so far all produced `centripetally' with their parts being made outside the entity and brought together toward the center.

The question of {\em `What is an individual?'} is a deep and difficult one for biology and artificial life, that we must return to again and again. Individuality is intimately intertwined with self-reproduction and evolution. 
The origin, maintenance and evolution of individuality for multicellular entities with differentiated cells has been and continues to be an active and fruitful area of
inquiry for  biology \cite{Buss,Michod}.
While Darwinian evolution describes dynamics of a population of individuals, it does not define what is meant by  `individual'.  Indeed, Darwinian evolution works whether the `individuals' are living organisms, viruses, programs in genetic programming, automata in evolutionary algorithms, or bitstrings in genetic algorithms.  
Moreover, the individuality of living organisms is not always discretely clear cut and often arises via the co-habitation of many `partners'; where we have the fertile concept of a {\em holobiont} comprising  a host organism and various microorganisms such as bacteria, viruses, and fungi that live within the host or just in a close association (of different individuals -- without any one identifiable as `host'), actively participating in its biology and sharing an intertwined fate with its lineage \cite{MargulisFester,Holobiont}.

\subsection{Von Neumann's Problem}\label{vNproblem}

In the 1940s, John von Neumann observed that generally, inanimate and human-made things we come across do not produce anything  as complex or more complex than themselves.
Von Neumann sought to resolve whether  self-reproduction requires ``life force'' or can  be explained by mechanistic / chemico-physical / logical means.
Is a rigorous  mathematical-physical  understanding of reproduction that we see in living things possible?

This is von Neumann's problem and he addressed it in two parts: 

\begin{enumerate}
    \item[vN-I.] How is it possible  for a mechanistic system to produce something as or more complex than itself?
    \item[vN-II.] How is it possible for complexity to increase over several generations of reproduction (as in biology)?
\end{enumerate}

The first one (vN-I) is solved explicitly in at least two fundamentally different ways by von Neumann in his book, most explicitly using cellular automata, while the second question (vN-II) is answered in principle by von Neumann's  ``Fifth Lecture'' \cite{vN} based on solution of the first question (vN-I), but with no implementation solving (vN-II) being given or described  in detail.
Von Neuman's conceptual solutions to the first part included both cellular automata and kinematic self-reproducer  models, with the allegory of a robot on a `pond' of components (warehouse) that is able to build a copy of itself by either of  two different mechanisms, each in itself constituting a solution to the first part of von Neumann's problem (vN-I): 
\begin{enumerate}
    \item[Mech. 1.] Self-examination: step-by-step build up of a copy (`offspring') by examining own structure. During this construction, at no time does there exist a separate  complete description of the entity.\footnote{One may consider this mechanism as reproduction in the absence of any separately identifiable component comprising a `genome'.}.
    \item[Mech. 2.] Build according to a description (program or `genome') from available components and copy the program to the `offspring'.
\end{enumerate}

Articulation of these two mechanisms was remarkable at the time, with the second one, Mech.~2, presaging the discovery of the role of DNA in life-as-we-know-it on earth: DNA structure as the molecular basis of heredity for life on earth was only discovered later in 1953  \cite{WatsonCrick1953}.  Indeed,
DNA replication `blindly copies' and  protein biosynthesis `executes' the code. So living things on earth satisfy Langton's criterion for self-reproduction (see Section~\ref{LangtonCriterion} below) and also von Neumann's (Section~\ref{VNcriterion}), employing Mech.~2. Alternatively, self-reproduction via self-examination (Mech.~1) could be achieved, e.g., by an entity doing by a {\it raster scan} in 2D or 3D of  itself and progressively printing a copy of itself as the scanning progresses,\footnote{Possibly this could be done at a molecular level in physical analogues of self-reproduction with 3D scanning and printing.} or by {\it template-matching} that copies the entire self, perhaps in two stages: producing a complementary strand and one complementary to the complement. Arguably, self-reproducing strands of RNA in a hypothesized pre-biotic RNA world are of this type \cite{RNAworld}. 

McMullin \cite{McMullin2000} offers an insightful discussion of von Neumann's pivotal contributions to understanding the evolutionary growth of complexity, the second part of von Neumann's problem (vN-II), as outlined in von Neumann's Fifth Lecture. Realization of von Neumann's solution, i.e., harnessing Darwinian evolution of self-reproducers, to this second problem in artificial systems  had to wait for novel mechanisms supporting inhertiable variation in H.~Sayama's evoloops \cite{Sayama1999a} and T.~S.~Ray's Tierra in the 1990s \cite{Ray1991}.

\subsection{Christopher Langton and His Views}
\label{LangtonCriterion}

To ensure non-triviality in the characterization of self-reproduction, Langton \cite{Langton1984} introduced the criterion presented below to exclude the trivial cases of self-replication described above (Section~\ref{TrivialSelfRep}).

His implemented solution in the Langton loop \cite{Langton1984} gives one answer to the first part of von Neumann's problem (vN-I), using the second strategy for self-reproduction described by von Neumann (Mech.~2), i.e., production of an offspring based on execution of instructions in a genome together with copying of that genome to the offspring.

To sum it up, his specific criterion for self-reproduction is:
\begin{itemize}
    \item {\bf Langton's Criterion for Non-trivial Self-Reproduction:} \\ To produce an offspring via self-reproduction,
    a system must blindly copy instructions to the offspring (genome), and these instructions must be executed to generate a  phenotype.
\end{itemize}

Following Langton's work, other researchers developed two directions:   First, simplifying the replicators as much as possible:  removal of one-side of the sheath around the  circulating genome of the loop, and later complete removal of the sheath \cite{Byl1989,ReggiaEtAl1993}.   While the steps in this simplification process preserve the dynamics of self-replication with the intention of meeting Langton's criteria, it is hard to say that the resulting minimal replicators are really non-trivial when we compare them with the trivial replicators in Section~\ref{TrivialSelfRep}. It seems that their complexity is in the eyes of the observer who knows the process that led to their invention.   Second,  adding complexity was pursued to give the self-replicating loops more computation power (stack automata capable of recognizing context free languages, or Turing complete computation); see the survey by Tempesti~\cite{Tempesti2012}. However, while the additional computation power allowed one to assert the replicators are computationally more complex than Langton's loop, this complexity  has  not  been used in any essential way  in the self-replication process.   It is an inessential ``add-on'' satisfying the mathematical criterion of higher complexity, but has not been used to achieve self-reproduction that supports inheritable variation nor has it served as a basis for evolution of further complexity.

While Langton's criterion nicely captures a prerequisite to one of the methods (Mech.~2) of solving von Neumann's problem (vN-I), it is silent about ensuring that complexity increase and an evolutionary process could be supported. That is, as much of the development following its introduction has shown, Langton's criterion is neither necessary nor sufficient for addressing the second aspect of von Neumann's problem (vN-II).

\subsection{Von Neumann's Criterion for Self-Reproduction}
\label{VNcriterion}

In {\em Theory of Self-Reproducing Automata} \cite[p.\ 86]{vN}, John von Neumann explicitly required a self-reproducer to have capacity for inheritable mutation.  We can formulate: 

\begin{itemize}
   \item {\bf Von Neumann's Criterion for Non-trivial Self-Reproduction:}
A self-reproducer must have capacity for inheritable mutation in addition to the ability to replicate itself.
\end{itemize}

\noindent
We use von Neumman's definition and criterion for self-reproduction in this review. Langton's loop is self-replicating and satisfies Langton's criterion, but not von Neumann's for a self-reproducing system.  Moreover, von Neumann's criterion is much more general in that it allows self-examination as an alternative mechanism for self-reproduction (Mech.~2),  also discussed by Laing \cite{Laing1977}.

Does von Neumann's criterion necessarily guarantee non-triviality?  It does guarantee that the reproducer could {\it in principle} be an individual in an evolutionary process. The fact that some populations undergoes  Darwinian evolution, say in a genetic algorithm or two-allele populations genetics model, does not necessarily entail much evolvability nor open-endedness. Similarly,  von Neumann's criterion might not necessarily lead to the evolution of complexity, nor even to any evolution at all  -- the context for Darwinian evolution may simply be lacking, as it is with von Neumann's own example~\cite{vN}, where there are no sources of variability (see~\cite{vN} for full details of von Neumann's universal constructor self-reproducer and \cite{Yinusa2011} for the variability and evolutionary considerations). 
However, von Neumann's criterion is a necessary requisite for the (Darwinian) evolution of complexity in any population of self-reproducers, and thus a good criterion for non-triviality. 
This more general notion of self-reproduction is the key to solving the second part of von Neumann's problem (vN-II) in full generality, as reviewed next.

\section{Natural Selection and Evolution in Cellular Automata}
\label{sec:evolution}

\subsection{Evoloops and Subsequent Studies through Mid-2010s}
\label{individual-evo}

Von Neumann's solutions to self-reproduction give potential mechanisms to harness the process of Darwinian evolution \cite{DarwinWallace1858}.  This is characterized as \textbf{a process on a 
 population of individuals} (self-reproducing in the case of nature), i.e., a population process with the properties  ({\bf Axioms for Darwinian Evolution}) \label{DarwinianEvolution} of:
 \begin{itemize} 
    \item \textbf{Heredity} - parents' traits are passed to offspring.
    \item \textbf{Variability} - offspring can differ from their parents.
    \item \textbf{Differing Fitness} - creating a selection in the population (which means a loss in diversity) depending at least in part on inherited traits.
    \item \textbf{Finite Lifetime} - a requisite for turn-over of generations.
    \item \textbf{Finite Space and Resources} - creating a struggle for existence.
\end{itemize}

In the case of living things, self-reproduction and natural selection are also exhibited:

\begin{itemize}
    \item \textbf{Self-Reproduction} - according to von Neumann's criterion (above). 
    \item \textbf{Intrinsic Fitness} - determining reproductive success emerges as a property of self-replication, and of the interaction among individuals and between an individual in its environments.
\end{itemize}

The above are necessary and sufficient criteria for life according to  Thomas S.~Ray and are exhibited by his Tierra system \cite{Ray1991}.
 These axioms entail a  struggle for existence, pressure for adaptive change in the population over generations,
and 
 account for the possibility of evolutionary change and  complexity increase of self-reproducers over time. 

In cellular automata, finite space and resources arise from the use of bounded  computational resources of the cellular automata space.\footnote{This is automatic for bounded grids with finite update times (i.e, larger some positive lower-bound duration $\delta$) including those with periodic boundary conditions (``wrap-around''), and in fact for all present day real-world computational devices, even though they might be set up to emulate an unbounded environment.  } Other of these aspects (Darwinian axioms) were successively addressed first by Langton's self-reproducing loops \cite{Langton1986} (yielding Heredity, using Mech.~2, and Finite Lifetime) and then work of Sayama who solved the problem of freeing-up resources for generations using structurally dissolvable variants of Langton's loop yielding a turn-over of generations \cite{Sayama1998}.  Then, Sayama improved the robustness of state-transition rules to collisions and interactions between loops that made Variability possible without catastrophic failure (inviability of loops) despite some changes to the heritable information within the loops \cite{Sayama1999a,Sayama1999b}. This combination yielded Differing Fitness, i.e., reproductive success depending in part on inherited characteristics and resulted in evoloops.

The original evoloop system \cite{Sayama1999a,Sayama1999b} was a 9-state 2D cellular automata model with von Neumann (5-cell) neighborhoods \cite{von Neumann1966},\footnote{The local synchronous update of the state at each location (cell) depends on the state of the cells in the 2D grid located above, below, left, and right of a given cell, and on the state of the cell itself.} derived from Langton's self-reproducing loops \cite{Langton1984}. Its state-transition function was revised thoroughly so that it would operate more robustly under a greater variety of local situations \cite{Sayama1999a} and that any undefined situation would generate a ``dissolving'' state that would propagate through contiguous active states and erase them from the space (structural dissolution) \cite{Sayama1998,Sayama1999a}. Turn-over of generations is made possible in this way by freeing up space which otherwise would become clogged with the debris and remnants from no longer active individuals (as in loop models prior to Sayama's), and the robustness frequently makes inheritable variation possible in loops still able to reproduce.

\begin{figure}[t]
\centering
\includegraphics[width=0.8\columnwidth]{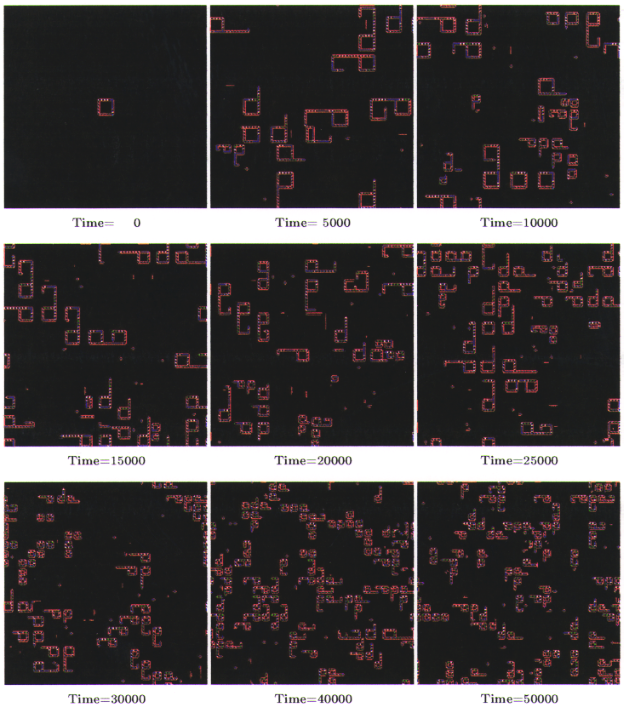}
\caption{Spontaneous evolution of evoloops (image from \cite{Sayama1999a}). Time indicates the number of updates applied.}
\label{evoloops}
\end{figure}

In the evoloop model, direct interactions (collisions) of loops caused irregular situations during their replication process, naturally inducing variations of their genomes. Such variations in the genomes of the surviving loops lead to inheritability of this variation in a lineage. Differing reproductive success emerged intrinsically without any global fitness function in evoloops. The loops that were better and faster and able to survive and reproduce did so in the physics of the cellular automata universe (Figure~\ref{evoloops}) \cite{Sayama1999a}. All state updates are local and the system includes no concept of ``individual'' or ``selection'' in its implementation. This was the first realization of von Neumann self-reproducers evolving by natural selection.  Remarkably, this was an instance of determinisic Darwinian evolution guaranteed to always play out the same way with the same initial conditions, with the trajectory of the evolving population depending on interactions between individuals for the generation of evolutionary novelty. The evoloop is also known to demonstrate substantial fault tolerance and, with slightly revised state transition function, abiogenesis from an initial configuration with no ancestor replicators \cite{Sayama1999b}.

The advent of evoloops triggered several subsequent studies that implemented important developments. Nehaniv \cite{Nehaniv2002} expanded the model to asynchronously updated cellular automata and showed that evolution can occur even without synchronous updating. Sayama \cite{Sayama2004} introduced self-protection behaviors of individuals and showed that it would help promote the diversity of species. Salzberg and Sayama \cite{Salzberg2004a} conducted detailed genetic sequencing of all the individual self-replicators that appeared in simulations and found that their genotypic/phenotypic diversities were much greater than originally thought and they continued to evolve for a long period of time after the loops' size reached the smallest level. Salzberg et al.\  \cite{Salzberg2004} also studied the evolutionary dynamics of evoloops in dynamic, hostile environments. 
A concise review of research on self-replication and evolution in cellular automata up to mid-2000s can be found in \cite{Salzberg2004b}.
Moreover, Oros and Nehaniv \cite{Oros2007} then presented a revised model called ``sexyloop'' in 2007 in which loops engaged in sexual reproduction by exchanging genetic information when colliding with each other; and in a later variant  in 2009, where the capacity for sex was made inheritable (rather than instrinsic in the physics of the update rules), they also showed that the capability of sexual reproduction can be evolutionarily maintained and impacts species diversity \cite{Oros2009}.

A related approach studied at about the same time was to achieve evolution of self-replicators via the shape-encoding mechanism. It was originally proposed by Morita and Imai \cite{Morita1996} for self-replication of various shapes in reversible cellular automata, but later adopted to promote spontaneous evolution of patterns through their spatial interactions \cite{Sayama2000, Suzuki2006}. 

Since then, this line of research experienced a relative dormancy of activities for about a decade, probably because the topics of interest in the artificial life community became diversified and shifted more toward evolutionary robotics, neuroevolution, swarm intelligence, agent interactions, and others. During this ``lost'' decade, there was not much progress made in self-reproducing and evolving cellular automata research, but some exceptions exist: In 2011, Yinusa and Nehaniv \cite{Yinusa2011} showed explicitly that mutations to the tape (genetic information encoding its construction) of von Neumann's self-reproducing automata can result in non-viable offspring or failure to complete the reproduction process (lethal deleterious mutations), as well as heritabilty of neutral mutation, and also in successful complexity increase in a lineage.  However, the modifications to the inheritable information on the tape were `genetically engineered' and not the result of a intrinsic process of variability generation. Another example is that, in 2013, Huang et al.\ \cite{Huang2013} proposed self-reproducing loops that adapt their shapes to local spatial constraints.

\subsection{Recent Trends Since Late 2010s: Open-ended Evolution and Continuous Cellular Automata}

Interestingly, for the last several years, there has been a resurgence of researchers' interest in spontaneous evolution of self-replicators in cellular automata. This is partly because of the Open-Ended Evolution (OEE) movement \cite{Taylor2016,packard2019a,packard2019b,stanley2019} that re-ignited the study of evolutionary dynamics within a dynamical system. The majority of OEE models used more conventional evolutionary computation approaches, but some researchers attempted more bottom-up, emergent evolutionary approaches using distributed dynamical systems like cellular automata \cite{Nichele2023}.

A few early examples of OEE research with cellular automata were published in 2017, one by Adams et al.\ \cite{Adams2017a,Adams2017b} and another by Andras \cite{Andras2017}. Adams et al.\ \cite{Adams2017a,Adams2017b} systematically investigated how to achieve OEE within 1D elementary cellular automata and showed that dynamic changes in environmental conditions (= transition rules) can exhibit OEE with unbounded evolution and innovation, with the state-dependent transition rule dynamics being most effective and most scalable. Similarly, Andras \cite{Andras2017} applied Bedau's metrics for open-ended evolutionary activity \cite{Bedau1992,Bedau1998} to cellular automata and showed that cellular automata worlds could exhibit OEE. Other relevant studies include Cisneros et al.'s work on detecting evolving structures in cellular automata dynamics \cite{Cisneros2019,Cisneros2023}. However, these studies did not consider evolutionary dynamics of non-trivial self-replicators/self-reproducers in Langton's or von Neumann's sense; they merely focused on complex spatio-temporal nonlinear dynamics of cellular automata configurations. Nonetheless, they helped re-ignite the research community's interest in cellular automata-based evolutionary models.

Recently, the success of continuous cellular automata models, such as Chan's Lenia \cite{Chan2019,Chan2020} and neural cellular automata \cite{Mordvintsev2020}, has attracted many researchers to explore how to create spontaneous evolutionary processes of diverse self-replicating patterns within a continuous cellular automata space. Most notably, Chan's Lenia world \cite{Chan2019,Chan2020} is an iconic example that beautifully demonstrated that a wide variety of `individuals' of virtual creatures with different phenotypes can stably exist, interact with each other and even replicate themselves within a distributed dynamical system like cellular automata. While Lenia was originally derived from Conway's Game of Life \cite{ConwayGoL1,ConwayGoL2}, the dynamic patterns emerging in Lenia were much more stable and much more biological-looking. They inspired many researchers (and hobbyists) to further explore possibilities of artificial life forms and their evolution in continuous cellular automata, and thus many studies of similar continuous cellular automata models followed \cite{kawaguchi2021,davis2022,suzuki2023,kojima2023}. Meanwhile, these models remained at the level of describing only self-replication, where evolution via variation and natural selection was not fully realized.

\begin{figure}
\centering
\includegraphics[width=\columnwidth]{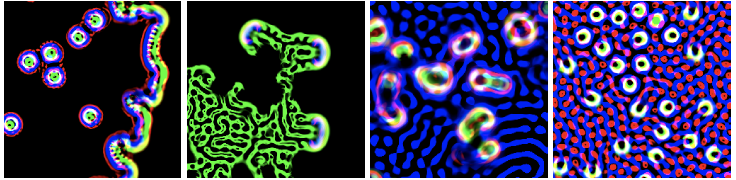}
\caption{Various patterns evolved within Evolutionary Lenia (images courtesy of Bert Chan; from \cite{Chan2023}).}
\label{evolutionary-lenia}
\end{figure}

The most recent development in this line of research is the attempt to make transition from self-replication to self-reproduction (and thus true evolution) in continuous cellular automata. For example, Sinapayen \cite{Sinapayen2023} trained neural cellular automata so that they replicate given organism patterns, and then demonstrated that the replicated patterns would deviate from the ancestor pattern over time, suggesting the possibility of genetic/phenotypic variation over multiple generations. Also, Plantec et al.\ \cite{Plantec2022} and Chan \cite{Chan2023} studied Lenia-based evolutionary systems in which model parameters were associated with each location and diffused over space (with mutations) so that multiple species and their interactions could be simulated simultaneously within a single simulation run (Figure \ref{evolutionary-lenia}). This is similar to the ``recipe'' propagation approach used in evolutionary Swarm Chemistry models \cite{Sayama2011} with great potential to generate a broad range of self-replicating spatio-temporal patterns automatically and efficiently.

\section{Future Challenges and Open Problems}\label{sec:Challenges}

These recent developments that utilize continuous cellular automata are very promising and expected to produce further advances in self-reproduction and evolution research. Meanwhile, many key questions still need to be addressed, and here we discuss only a few. 

First, there are many concrete challenges and open problems in self-reproduction and evolution for von Neumann's self-reproducing universal constructors \cite{vN,Yinusa2011}. One group concerns extensions of von Neumann's original 29-state cellular automaton model of self-reproduction (using strategy Mech.~2): 
\begin{enumerate}
\item[I.1.] Extend von Neumann's self-reproduction system so that multiple descendants can be active at once, e.g.,
the tape specifies two offspring.
\item[I.2.] Introduce robustness to collisions and to noise to make reproduction robust to these (similar to
evoloop).
\item[I.3.] Demonstrate Darwinian evolution in a population of such variant von Neumman universal constructor self-reproducers, and
study evolutionary trajectories and scenarios.  (Notice that T.S.~Ray's definition of life would be satisfied: Darwinian evolution with self-reproduction and natural selection).
\item[I.4.]  Introduce sex into von Neumann reproducers.
\item[I.5.] Integrate self-repair and self-maintenance capacity into von Neumann self-(re)producers towards to achievement of autopoiesis.\footnote{To achieve this autopoiesis, the concise characterization of Varela \cite{Varela2000} could be implemented for self-reproducers: (1) semipermeable boundary, (2) a network of reactions, and (3) interdependence: the network of  reactions is regenerated by the existence of the same boundary, so the boundary and the network depend on one another \cite{Varela2000}. } 
This is an additional essential criteria for life, a milestone so far not achieved in any artificial self-reproducers.\footnote{Autopoietic systems are arguably individuals (see Sec.~\ref{Individuals}) {\em from their own side}  and thus are natural candidates for the basic entities of the population required by the axioms of Darwinian evolutionary process by natural selection (Sec.~\ref{DarwinianEvolution}).}
\end{enumerate}

A further set of hard challenges refers to self-reproducers going beyond the methods of the original von Neumann self-reproducer (that was completed by A.W. Burks and others):
\begin{enumerate}
\item[II.1.] Implement a universal self-reproducing constructor that uses self-examination to reproduce, either in cellular automata, or
physically. 
\item[II.2.] Realize and/or transfer all of the above in continuous cellular automata.
\end{enumerate}

Second, nearly all the above evolutionary systems built within spatially distributed media exhibited the eventual dominance by one or a few most successful species in the long run, and it is still unclear what kind of generalizable principles or mechanisms are available to prevent the evolving ecosystem of self-replicators from falling into such pseudo-equilibrium states. It has been suggested that dynamic environments \cite{Salzberg2004,Sayama2011,Adams2017a,Adams2017b} are the key to addressing this issue, although they may not work for indefinitely long terms \cite{Sayama2018}. Overcoming this empirical limitation is a necessary and critical step towards implementing open-ended self-reproducing and evolving systems within cellular automata and other similar spatially distributed computational media.

Third, the recent approach to assign model parameters to local regions/agents deliberately avoids explicit representation of such genetic information in the space, in contrast to real biological systems and earlier self-replication models \cite{von Neumann1966,Langton1984,Sayama1998,Sayama1999a,Sayama1999b} where genetic instructions were explicitly represented in space. It is not well understood how these two approaches differ regarding the open-endedness and creativity of their evolution. They also differ significantly with regard to the computational capability built in the environment and in the ``laws of physics''. Gaining an in-depth understanding of the pros and cons of these two approaches will greatly inform us about the design choices for future models of self-reproduction and evolution in cellular automata.

Fourth, all the models reviewed so far relied on either logically designed mechanisms written in discrete state-transition rules or dynamically generated quasi-stable patterns in continuous space, but neither would capture the autopoietic nature of real biological systems. There is another major body of literature on computational autopoietic models \cite{McMullin2004,Nehaniv2005,Ikegami2008,Suzuki2009,Sirmai2011,Sirmai2013,Kliska2024,Lawrence2024}, and some of them even demonstrated self-replication of autopoietic structures \cite{Sirmai2013,Lawrence2024}. However, it remains unclear how one could integrate autopoietic dynamics into existing cellular automata models of evolving self-reproducing patterns. This is a potentially promising yet largely under-explored subject of study on cellular automata-based self-reproduction and evolution, which might eventually lead to a completely brand-new approach.

Fifth and finally, the relationship between the evolutionary dynamics of those self-replicators and the ``intelligence'' therein is worth further quantitative investigation. Intelligence is often associated with computational universality and critical behavior, and therefore, it may be characterized by the incompressibility/irreducibility of spatio-temporal dynamics \cite{Cisneros2019,Cisneros2023}. It would be quite interesting to quantify how much compressibility/reducibility the spatio-temporal dynamics of evolving self-replicators would have, and how it would change over time in the course of their evolution. If one could create evolutionary cellular automata that become increasingly harder to simulate over time, that would indicate the increasing complexity (= computational capability, ``intelligence'') of evolving entities within spatially distributed media. This would be a direct, concrete, measurable demonstration of the very original motivating vision posed by von Neumann \cite{von Neumann1966}.


\end{document}